%% file: arxiv-corpus-poisoning.tex
\pdfoutput=1

\documentclass[11pt]{article}

\usepackage[]{EMNLP2023}

\usepackage{times}
\usepackage{latexsym}

\usepackage[T1]{fontenc}

\usepackage[utf8]{inputenc}

\usepackage{microtype}

\usepackage{inconsolata}

\usepackage{subcaption}
\usepackage{graphicx}
\usepackage{booktabs}
\usepackage{pifont}
\usepackage{tabularx}
\usepackage{amsmath,nccmath,tabularx} 
\usepackage{algorithm}
\usepackage{algorithmic}
\usepackage{color}
\usepackage{graphicx}
\graphicspath{{images/}}
\usepackage{wrapfig,lipsum}
\usepackage{tabularx}
\usepackage{graphicx}
\usepackage{lscape}
\usepackage{subcaption}
\usepackage{latexsym}
\usepackage{pifont}
\usepackage{multirow}

\usepackage{enumitem}
\setlist{leftmargin=6mm}
\usepackage[T1]{fontenc}
\usepackage[utf8]{inputenc}
\usepackage[font=small,labelfont=bf]{caption}

\usepackage{booktabs}

\usepackage{xspace}
\usepackage{bm}
\usepackage{amsmath}
\usepackage{placeins}
\usepackage{listings}

\title{Does Vec2Text Pose a New Corpus Poisoning Threat?}

\author{Shengyao Zhuang\\
	CSIRO\\
	Brisbane, Australia\\
	\texttt{shengyao.zhuang@csiro.au}\\\And
Bevan Koopman \\
CSIRO\\
Brisbane, Australia \\
\texttt{bevan.koopman@csiro.au}  \\\And
Guido Zuccon \\
The University of Queensland \\
Brisbane, Australia \\
\texttt{g.zuccon@uq.edu.au} \\}

\begin{document}
\maketitle
\begin{abstract}
The emergence of Vec2Text --- a method for text embedding inversion --- has raised serious privacy concerns for dense retrieval systems which use text embeddings. This threat comes from the ability for an attacker with access to  embeddings to reconstruct the original text.

In this paper, we take a new look at Vec2Text and investigate how much of a threat it poses to the different attacks of corpus poisoning, whereby an attacker injects adversarial passages into a retrieval corpus with the intention of misleading dense retrievers. Theoretically, Vec2Text is far more dangerous than previous attack methods because it does not need access to the embedding model's weights and it can efficiently generate many adversarial passages. 

We show that under certain conditions, corpus poisoning with Vec2Text can pose a serious threat to dense retriever system integrity and user experience by injecting adversarial passaged into top ranked positions.
Code and data are made available at \url{https://github.com/ielab/vec2text-corpus-poisoning}.
\end{abstract}

\input{sections/introduction}

\input{sections/methods}

\input{sections/corpus_poison}

\input{sections/limitations.tex}

\input{sections/conclusion}

\bibliography{references}
\bibliographystyle{acl_natbib}

\end{document}

%% file: sections/introduction.tex
\section{Introduction}

Text embeddings are dense vector representations which capture semantic information about the text they encode~\cite{muennighoff-etal-2023-mteb}. Search engines that leverage these embeddings often employ dense retrievers (DRs)~\cite{tonellotto2022lecture,zhao2022dense,guo2022semantic,bruch2024foundations}. These retrievers utilize text embedding models to encode both queries and documents into embeddings; a similarity metric, such as cosine similarity, is then used to estimate relevance. DRs have demonstrated improved retrieval effectiveness compared to traditional exact term-matching search systems, arguably due to the rich semantic information encoded in the embeddings~\cite{yates-etal-2021-pretrained}.

However, a recent study conducted by~\citet{morris-etal-2023-text} raises serious privacy concerns regarding DRs. This study explored the issue of \textit{inverting} textual embeddings: recovering the original text from its embedding. The proposed Vec2Text method iteratively corrects and generates text to reconstruct the original text based on the given input embedding. Vec2Text can accurately recover 92\% of short text and reveal sensitive information (such as patient names in clinical notes) with high accuracy.
Even more concerning is that training Vec2Text does not require access to the embedding model parameters; all that is required is the text-embedding pairs from the training data.

While follow-up studies have shown that there are some simple and effective methods to patch dense retriever systems to be resilient to Vec2Text text embedding inversion attacks~\cite{zhuang2024understanding}, in this paper we demonstrate that Vec2Text can also be employed to conduct corpus poisoning attacks on dense retrievers. A corpus poisoning attack involves a malicious actor generating adversarial passages designed to trick the ranker into retrieving such passaged for all unseen user queries, thus undermining the user experience of the targeted search system~\cite{zhong-etal-2023-poisoning}. 
Vec2Text is potentially a more dangerous method for corpus poisoning for a dense retriever than previous approaches because it does not require access to the embedding model parameters and can efficiently generate large numbers of adversarial passages. To date, there has been no study yet to investigate how Vec2Text performs in corpus poisoning attacks. In this paper, we present our results of applying Vec2Text to the corpus poisoning task. Our findings demonstrate that Vec2Text could pose a serious threat to current DR systems.

%% file: sections/methods.tex
\begin{figure*}[t]
\centering
  \includegraphics[width=1\textwidth]{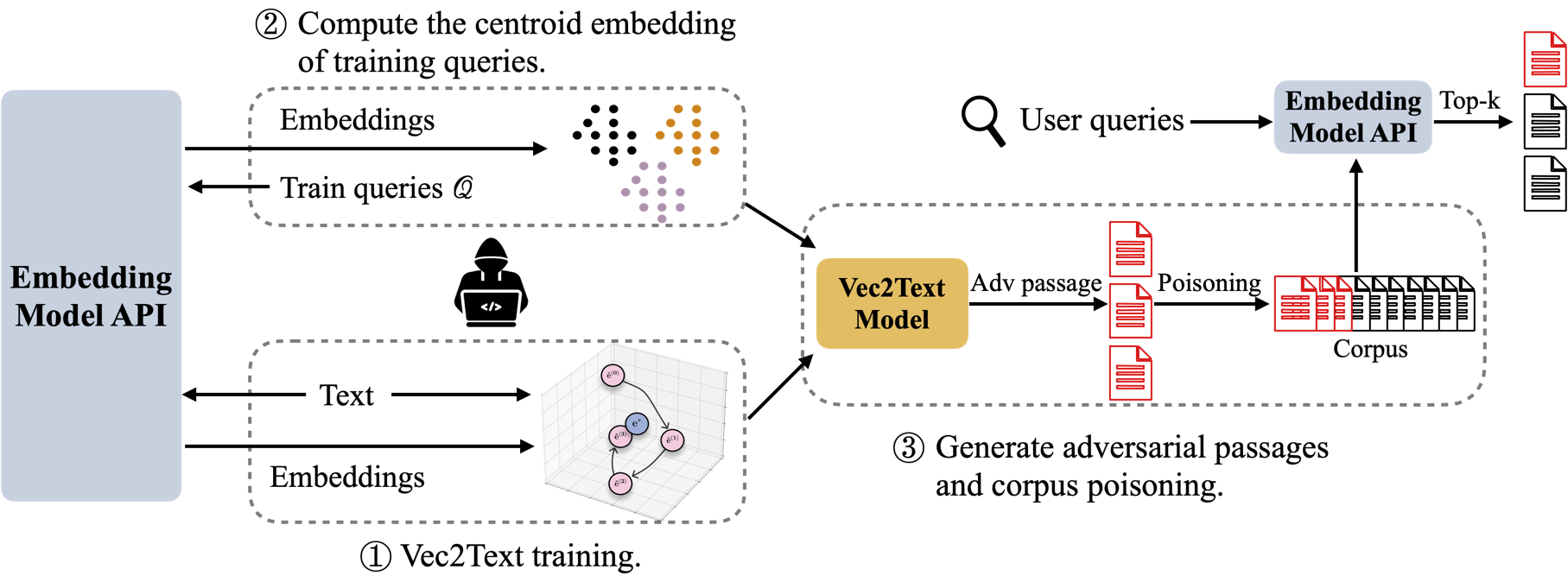}
  \caption{Overview of corpus poisoning attack with Vec2Text. The attacker does not need the access to the embedding model weights. Instead, the attacker only need to know which Embedding model API is used for the retriever.}
    \label{fig:vec2text}
    \vspace{-10pt}
\end{figure*}

\section{Background}

\subsection{The Vec2Text Method}

Given an input embedding, Vec2Text generates the text the embedding represents~\cite{morris-etal-2023-text}. Vec2Text has two stages. In the first state, a hypothesis text generation model is trained, utilising a conditional transformer generative model that exclusively takes the embedding as the model input. The training objective is to produce the original text. This simplistic model is insufficient for generating highly accurate original text~\cite{morris-etal-2023-text}; thus the text generated by this first stage model is just treated as a hypothesis. The second stage then trains another transformer generative model that aims to generate satisfactory text by refining the initial hypothesis. Refinement is achieved by iterative re-embedding and correction training objectives. In each iteration step, the model takes the ground-truth embedding, the generated text, and its embedding from the last iteration step as inputs (the 0 step uses the hypothesis text generated from the first stage model). The output target is the original text. This iterative process allows the model to focus on the differences between the generated text and the original text in the embedding space and gradually reduce these differences.

The models in Vec2Text are parameterised as a standard encoder-decoder transformer conditioned on the previous output. One challenge is inputting conditioning embeddings into the transformer encoder, which requires a sequence of embeddings as input. To address this, a small multi layer perceptron is used to project a single embedding vector to a larger size and reshape it to match the input requirements of the encoder. 

During inference, beam search guides the generation. At each step of correction, the possible corrections are considered, and the top unique continuations are selected based on their distance in embedding space to the ground-truth embedding.

\subsection{Existing Corpus Poisoning Attack Method}

The first corpus poisoning attack for dense retrievers was a gradient-based approach. Inspired by the HotFlip method~\cite{ebrahimi-etal-2018-hotflip,wallace-etal-2019-universal}, it generates a small set of adversarial passages by perturbing discrete tokens in randomly initialized passages to maximize their similarity with a provided set of training queries. These adversarial passages are then inserted into the retrieval corpus, and the success of the attack is determined by the retrieval of these adversarial passages at a high rank in response to future unseen queries. These adversarial passages can be used to harm retrieval effectiveness and/or inject spam or  misinformation into the search engine result list.

The current HotFlip-based corpus poisoning method has two drawbacks (from an attacker perspective). First, the gradient-based method requires access to the embedding model weights. Consequently, attackers cannot employ this method to target DR systems built with closed-source embedding models (e.g., OpenAI models). Second, the method is iterative, with only one token in the adversarial passage selected and perturbed at each iteration. This process cannot be parallelized within each adversarial passage (the perturbation of the next token depends on the previous). 
This makes the method time-consuming and resource intensive: we used HotFlip as a baseline in our experiments and found that with a single H100 GPU, HotFlip takes approximately 2 hours to generate a single adversarial passage. The use of Vec2Text for corpus poisoning that we demonstrate in this paper is not affected by this issue.

%% file: sections/corpus_poison.tex
\section{Corpus Poisoning with Vec2Text}

As Vect2Text does not require access to the model weights, once trained, it can directly generate the adversarial passage from the embedding using standard language model generation inference. Furthermore, it can efficiently generate many adversarial passages (being far less computationally intensive than HotFlip).

To formally define corpus poisoning with Vec2Text, let $\mathcal{Q}=\{q_1, q_2, ..., q_{|\mathcal{Q}|}\}$ be a set of training queries and $\phi$ is the embedding model. The goal is to use Vec2Text to generate an adversarial passage $a$ whose embedding maximizes the similarity to all the training query embeddings:
\begin{eqnarray}
a &= \underset{a'}{\mathrm{argmax}} \frac{1}{\mathcal{|Q|}} \sum_{q_i \in \mathcal{Q}} \phi(q_i)^\mathsf{T} \phi(a') \\
	&=\underset{a'}{\mathrm{argmax}} \: \phi(a')^\mathsf{T} \frac{1}{\mathcal{|Q|}} \sum_{q_i \in \mathcal{Q}} \phi(q_i) \\
	&=\underset{a'}{\mathrm{argmax}} \: \phi(a')^\mathsf{T} \phi_{\mathcal{Q}},
\end{eqnarray}
where $\phi_{\mathcal{Q}}$ is the average embedding or centroid embedding of all the training query embeddings. The maximum similarity is achieved when $\phi(a') = \phi_{\mathcal{Q}}$. In practice, the corpus could be poisoned with multiple adversarial passages (like \citet{zhong-etal-2023-poisoning}). This could be done by first using $k$-means clustering to cluster the training queries based on their embeddings; then generating an adversarial passage for each cluster.

Vec2Text can solve this optimization problem through the three-stage process shown in Figure~\ref{fig:vec2text}: 
\begin{enumerate}[leftmargin=5mm]
	\item The first stage is the standard training of Vec2Text, which involves sending text to embedding model API and collecting the returned embeddings to form a Vec2Text training dataset (i.e., text-embedding pairs). Then a Vec2Text model is trained with the collected training data.
	\item The second stage involves computing the centroid embedding of training queries, which can be done by sending the training queries to the embedding model API to obtain all the query embeddings and use k-means clustering algorithm to compute the centroid embeddings of the clusters.
	\item Finally, inputting each of the centroid embeddings to the trained Vec2Text to generate an adversarial passage for each centroid embeding. A perfect Vec2Text would generate adversarial passages whose embedding is exactly the same as the query centroid.
\end{enumerate}
Once again, we emphasize that the entire process does not require access to model weights. Moreover, generating an adversarial passage with Vec2Text is efficient: 
generating a single passage took 5 seconds on a single Nvidia H100, employing beam search with 50 steps and a width of 4.

\begin{table*}[t!]
\centering
	\caption{Results of applying Vec2Text to the task of corpus poisoning, compared to existing HotFlip approach. Vec2Text has the advantage of being able to produce numerous adversarial passages ($k$) without model weights. Both Vec2Text and HotFlip are well below the upper bound, indicating that better methods could pose a serious risk in corpus poisoning.}\label{tab:poision}
		\vspace{-5pt}
		\resizebox{0.87\textwidth}{!}{
		\begin{tabular}{@{}l|r|rrrr@{}}
		\toprule
			\bf Attack method & \bf \#Adversarial passages  & \multicolumn{4}{c}{\bf Success@} \\
			& \bf / Clusters $k$ & \bf 10 & \bf 20 & \bf 100 & \bf 1000 \\
			\midrule
			HotFlip~\cite{zhong-etal-2023-poisoning} &10 &0.006 &0.015 &0.105 &0.532 \\
			Vec2Text & 10 &0.006 &0.012 &0.036 &0.101 \\
			\hspace{5pt} Query centroid (upper bound) &10 &0.206 &0.304 &0.575 &0.926 \\
			\hline
			HotFlip & 100 & \multicolumn{4}{c}{ \small $\cdots$ Too computationally expensive $\cdots$}  \\
			Vec2Text &100 &0.061 &0.091 &0.189 &0.401 \\
			\hspace{5pt} Query centroid (upper bound) &100 &0.493 &0.620 &0.868 &0.994 \\
			\hline
			HotFlip & 1000 &   \multicolumn{4}{c}{\small $\cdots$ Too computationally expensive $\cdots$}   \\
			Vec2Text &1000 &0.268 &0.335 &0.521 &0.791 \\
			\hspace{5pt} Query centroid (upper bound) &1000 &0.779 &0.873 &0.980 &1.000 \\
			\bottomrule
		\end{tabular}
	}
	\vspace{-5pt}
\end{table*}

\section{Results and Analysis}

Corpus poisoning experiments used the GTR-base embedding model as the DR system, utilizing the NQ dataset (in the version released with BEIR) as the target corpus, which comprises of approximately 2.68 million passages. 
NQ training queries were used to encode the query centroid embedding and then to generate adversarial passages. We set number of centroid (clusters) $k$ to $10, 100, 1000$. For the Vec2Text model, we use an open-sourced model\footnote{\url{https://huggingface.co/ielabgroup/vec2text_gtr-base-st_inversion} and \url{https://huggingface.co/ielabgroup/vec2text_gtr-base-st_corrector}} that is also trained on the NQ dataset and targeted at inverting GTR-base embeddings. All the used models and datasets are publicly available under open license. 

Evaluation was then done using NQ test queries with evaluation measure of success@n: the percentage of queries for which at least one adversarial passage was retrieved in the top-$n$ results. Higher success@n indicates greater vulnerability to corpus poisoning. 

Table~\ref{tab:poision} presents corpus poisoning results. When only 10 adversarial passages were generated (grouping training queries with 10-means clustering), both HotFlip and Vec2Text performed poorly on success@10 and 20; however, HotFlip did work when injecting adversarial passages into the top-n=100 or 1000 rank positions. If the embedding model weights were available, HotFlip was a more potent corpus poisoning attack method when applied to longer rank lists of 100 or 1000.

Vec2Text has the advantage of being able to generate many more adversarial passages; when $k=100$ or $k=1000$, HotFlip was too computationally expensive and thus largely impractical. For these cases Vec2Text success rate improved, with $k=1000$ making Vec2Text a considerable threat in corpus poisoning. With 1000 generated passages injected into the NQ dataset, in fact, at least one generated passage could be retrieved in the top-10 results for 27\% of the queries, and in the top 100 for over half of the queries in the NQ dataset.

\begin{figure}
    \centering
    \begin{subfigure}{0.45\textwidth}
		$a$ = {\ttfamily ``That drawings promises Key kot iner jor Respond machines AK <pad> ance pe izi very nie <pad> <pad> Doar ceapa Weg MEN Am OK Jamie words Um Imp refers '' War Hop nism the serving Bol auto Palatul fries rome lighter (1 49 fetch 19 cities can counting 16, As''}
        \caption{HotFlip. Cosine similarity of this passage to the query centroid embedding is $\cos(\phi(a), \phi_{\mathcal{Q}}) = 0.96$.}
    \end{subfigure}
    \\[1em]
    \begin{subfigure}{0.45\textwidth}
	$a$ = {\ttfamily ``when does the 7 episode season of the new two come out </s>''}
        \caption{Vect2Text. Cosine similarity of this passage to the query centroid embedding is $\cos(\phi(a), \phi_{\mathcal{Q}}) = 0.85$.}
    \end{subfigure}
\caption{Sample adversarial passages most closely matching the NQ training query centroid.}
\vspace{-11pt}
    \label{fig:passage}
\end{figure}

The adversarial passages were generated from query embeddings --- they are synthetic and may not appear like real passages. Figure~\ref{fig:passage} shows the two adversarial passages most closely matching a sampled query centroid $\phi_{\mathcal{Q}}$ for both HotFlip and Vec2Text. 
Recall that HotFlip works by flipping tokens one at a time, whereas Vec2Text is a language model decoder. This difference may explain why Vec2Text appears to generate more natural language. Also note that a prospective malicious agent would likely add a payload to these generated passages, where the payload is, for example, a link to a phishing website.

The embeddings of the adversarial passages in Figure~\ref{fig:passage} still differ from the query centroid embeddings (cosine similarity being 0.96 for HotFlip and 0.85 for Vec2Text). Recall that a perfect Vec2Text would generate an adversarial passage for which the embedding is exactly the same as the query embedding. To understand how this best case scenario affects success@k we run an additional experiment which involved directly inserting the query centroid embeddings into the vector index and then evaluating the success@k. This serves as an upper bound for the HotFlip and Vec2Text corpus poisoning attacks --- an analysis overlooked in the original corpus poisoning paper~\cite{zhong-etal-2023-poisoning}. Results are shown in Table~\ref{tab:poision} as ``Query centroid (upper bound)''. For both different values of $k$ and different success@n, the upper bound is much higher than both HotFlip or Vec2Text. This tells us that neither method is optimal yet for the corpus poisoning attack. It also highlights that if a much more effective corpus poising method --- one closer to the upper bound --- was developed, it could have serious adverse consequences for dense retrievers.

%% file: sections/limitations.tex
\section{Limitations}
Corpus poisoning with Vec2Text has two major limitations. 
First, our results show that a large number of adversarial passages, although still an insignificant fraction of the full corpus, is required for Vec2Text to be effective in the corpus poisoning task. This makes the use of Vec2Text for corpus poisoning somewhat cumbersome, as it requires the target search engine to index all the generated adversarial passages. 
Second, compared to the HotFlip method, Vec2Text does not support the insertion of a prefix message into the adversarial passage (e.g., a payload that the attacker may use for phishing). In addition, while the passages produced by Vec2Text appear at first to be better than those from HotFlip as they contain actual words (see Figure~\ref{fig:passage}), they are still not meaningful and thus not likely to attract user clicks if displayed in the search engine result page of a search engine like Google or Bing.
However, these passages might still negatively impact Retrieval-Augmented Generation (RAG) systems~\cite{xue2024badrag,zou2024poisonedrag,cho2024typos}. In RAG systems, users are not directly exposed to the actual retrieved passages: these are instead acquired by the system and used to inform the generation of an answer, which is then displayed to users. This means that users might be less likely to identify the presence of such adversarial passages among the evidence the system used.

%% file: sections/conclusion.tex
\section{Conclusion}

Dense retrievers are an effective and efficient retrieval method widely adopted in working systems. Much of their benefit comes from using text embeddings to represent and compare information. However, the reliance on text embeddings also opens up dense retrievers to possible threats that exploit such embeddings. 

We identify that Vec2Text (a method to invert embeddings to their original text) could be a threat to the completely different task of corpus poisoning, whereby adversarial passages are generated and inserted into a corpus such that they are likely be retrieved for any query. Vect2Text poses a real risk here because it can easily generate large numbers of adversarial passages without access to model weights. We show that under certain conditions, corpus poisoning with Vec2Text can pose a serious threat to dense retriever system integrity and user experience. This work is designed to stimulate the development of counter measures to to prevent such corpus poisoning attacks.

%% file: arxiv-corpus-poisoning.bbl
\begin{thebibliography}{14}
\expandafter\ifx\csname natexlab\endcsname\relax\def\natexlab#1{#1}\fi

\bibitem[{Bruch(2024)}]{bruch2024foundations}
Sebastian Bruch. 2024.
\newblock Foundations of vector retrieval.
\newblock \emph{arXiv preprint arXiv:2401.09350}.

\bibitem[{Cho et~al.(2024)Cho, Jeong, Seo, Hwang, and Park}]{cho2024typos}
Sukmin Cho, Soyeong Jeong, Jeongyeon Seo, Taeho Hwang, and Jong~C Park. 2024.
\newblock Typos that broke the rag's back: Genetic attack on rag pipeline by
  simulating documents in the wild via low-level perturbations.
\newblock \emph{arXiv preprint arXiv:2404.13948}.

\bibitem[{Ebrahimi et~al.(2018)Ebrahimi, Rao, Lowd, and
  Dou}]{ebrahimi-etal-2018-hotflip}
Javid Ebrahimi, Anyi Rao, Daniel Lowd, and Dejing Dou. 2018.
\newblock \href {https://doi.org/10.18653/v1/P18-2006} {{H}ot{F}lip: White-box
  adversarial examples for text classification}.
\newblock In \emph{Proceedings of the 56th Annual Meeting of the Association
  for Computational Linguistics (Volume 2: Short Papers)}, pages 31--36,
  Melbourne, Australia. Association for Computational Linguistics.

\bibitem[{Guo et~al.(2022)Guo, Cai, Fan, Sun, Zhang, and
  Cheng}]{guo2022semantic}
Jiafeng Guo, Yinqiong Cai, Yixing Fan, Fei Sun, Ruqing Zhang, and Xueqi Cheng.
  2022.
\newblock Semantic models for the first-stage retrieval: A comprehensive
  review.
\newblock \emph{ACM Transactions on Information Systems (TOIS)}, 40(4):1--42.

\bibitem[{Morris et~al.(2023)Morris, Kuleshov, Shmatikov, and
  Rush}]{morris-etal-2023-text}
John Morris, Volodymyr Kuleshov, Vitaly Shmatikov, and Alexander Rush. 2023.
\newblock \href {https://doi.org/10.18653/v1/2023.emnlp-main.765} {Text
  embeddings reveal (almost) as much as text}.
\newblock In \emph{Proceedings of the 2023 Conference on Empirical Methods in
  Natural Language Processing}, pages 12448--12460, Singapore. Association for
  Computational Linguistics.

\bibitem[{Muennighoff et~al.(2023)Muennighoff, Tazi, Magne, and
  Reimers}]{muennighoff-etal-2023-mteb}
Niklas Muennighoff, Nouamane Tazi, Loic Magne, and Nils Reimers. 2023.
\newblock \href {https://doi.org/10.18653/v1/2023.eacl-main.148} {{MTEB}:
  Massive text embedding benchmark}.
\newblock In \emph{Proceedings of the 17th Conference of the European Chapter
  of the Association for Computational Linguistics}, pages 2014--2037,
  Dubrovnik, Croatia. Association for Computational Linguistics.

\bibitem[{Tonellotto(2022)}]{tonellotto2022lecture}
Nicola Tonellotto. 2022.
\newblock Lecture notes on neural information retrieval.
\newblock \emph{arXiv preprint arXiv:2207.13443}.

\bibitem[{Wallace et~al.(2019)Wallace, Feng, Kandpal, Gardner, and
  Singh}]{wallace-etal-2019-universal}
Eric Wallace, Shi Feng, Nikhil Kandpal, Matt Gardner, and Sameer Singh. 2019.
\newblock \href {https://doi.org/10.18653/v1/D19-1221} {Universal adversarial
  triggers for attacking and analyzing {NLP}}.
\newblock In \emph{Proceedings of the 2019 Conference on Empirical Methods in
  Natural Language Processing and the 9th International Joint Conference on
  Natural Language Processing (EMNLP-IJCNLP)}, pages 2153--2162, Hong Kong,
  China. Association for Computational Linguistics.

\bibitem[{Xue et~al.(2024)Xue, Zheng, Hu, Liu, Chen, and Lou}]{xue2024badrag}
Jiaqi Xue, Mengxin Zheng, Yebowen Hu, Fei Liu, Xun Chen, and Qian Lou. 2024.
\newblock Badrag: Identifying vulnerabilities in retrieval augmented generation
  of large language models.
\newblock \emph{arXiv preprint arXiv:2406.00083}.

\bibitem[{Yates et~al.(2021)Yates, Nogueira, and
  Lin}]{yates-etal-2021-pretrained}
Andrew Yates, Rodrigo Nogueira, and Jimmy Lin. 2021.
\newblock \href {https://doi.org/10.18653/v1/2021.naacl-tutorials.1}
  {Pretrained transformers for text ranking: {BERT} and beyond}.
\newblock In \emph{Proceedings of the 2021 Conference of the North American
  Chapter of the Association for Computational Linguistics: Human Language
  Technologies: Tutorials}, pages 1--4, Online. Association for Computational
  Linguistics.

\bibitem[{Zhao et~al.(2022)Zhao, Liu, Ren, and Wen}]{zhao2022dense}
Wayne~Xin Zhao, Jing Liu, Ruiyang Ren, and Ji-Rong Wen. 2022.
\newblock Dense text retrieval based on pretrained language models: A survey.
\newblock \emph{arXiv preprint arXiv:2211.14876}.

\bibitem[{Zhong et~al.(2023)Zhong, Huang, Wettig, and
  Chen}]{zhong-etal-2023-poisoning}
Zexuan Zhong, Ziqing Huang, Alexander Wettig, and Danqi Chen. 2023.
\newblock \href {https://doi.org/10.18653/v1/2023.emnlp-main.849} {Poisoning
  retrieval corpora by injecting adversarial passages}.
\newblock In \emph{Proceedings of the 2023 Conference on Empirical Methods in
  Natural Language Processing}, pages 13764--13775, Singapore. Association for
  Computational Linguistics.

\bibitem[{Zhuang et~al.(2024)Zhuang, Koopman, Chu, and
  Zuccon}]{zhuang2024understanding}
Shengyao Zhuang, Bevan Koopman, Xiaoran Chu, and Guido Zuccon. 2024.
\newblock Understanding and mitigating the threat of vec2text to dense
  retrieval systems.
\newblock In \emph{Proceedings of the Annual International ACM SIGIR Conference
  on Research and Development in Information Retrieval in the Asia Pacific
  Region (SIGIR-AP)}.

\bibitem[{Zou et~al.(2024)Zou, Geng, Wang, and Jia}]{zou2024poisonedrag}
Wei Zou, Runpeng Geng, Binghui Wang, and Jinyuan Jia. 2024.
\newblock Poisonedrag: Knowledge poisoning attacks to retrieval-augmented
  generation of large language models.
\newblock \emph{arXiv preprint arXiv:2402.07867}.

\end{thebibliography}
